# Expected velocity anomaly for the Earth flyby of Juno spacecraft on October 9, 2013

H. J. Busack

Wulfsdorfer Weg 89, 23560 Lübeck, Germany
September 25, 2013

Abstract

The so-called flyby anomaly is a yet unexplainable velocity jump measured at several Earth flybys of spacecraft. Known physical effects could be excluded as source of this anomaly. In order to model a possible new physical effect, empirical equations were proposed by Busack (2007) and Anderson et al. (2007), which gave quite good description of all measured anomalies. Some theories were suggested deriving the Anderson formula or a similar one. The recent two Earth flybys of the spacecraft Rosetta showed no measurable anomaly, although the Anderson formula predicted distinct effects for both flybys. The Busack formula predicted the null results, so the notion of a possibly correct formula or of an error of the older measuring software was supported. The forthcoming Earth flyby of Juno gives a good opportunity to decide this question or give rise to enhanced theory, because the orbit parameters are very similar to earlier flybys with notable effects. In this article, the flyby anomaly according to the Busack equation will be predicted to be about -7mm/s in contrast to the value after the Anderson equation and similar ones with distinct positive value of the order of +6mm/s.

## 1 Introduction

Several Earth flybys of diverse spacecraft have shown a distinct velocity jump past closest approach (CA) to Earth when analyzing the Doppler and range measurements with the common orbit determination programs (ODP). According to several studies [2,3,8,10], this so-called flyby anomaly cannot be declared by unconsidered known physical effects. So there are but three possible sources for this effect: Failure of the OPD, measurement or compensation errors, or a real physical effect, not covered by conventional physics. For the latter possibility, diverse theories were suggested [11,12,13,14,15], but none of them were able to derive all measured anomaly values within the claimed error bar. Additionally, empirical equations were presented by Busack [4] and Anderson [7] and several theories were proposed [9,12,13,15] claiming to derive the Anderson formula or a similar one. After the latest two Earth flybys (Rosetta spacecraft, 2007 and 2009) with no detectable anomaly, it was clear that the Anderson formula had failed, as well as all theories deriving this or similar formula. Only the Busack equation showed the correct values. This equation models empirically an asymmetric gravitation field with definite special orientation. The asymmetry is determined by the direction and speed of the Earth´s movement within a hypothetical "gravitational rest frame", described by 3 parameters. Shape and strength of the asymmetry is determined by additional three parameters. Of course, the successful prediction of the two recent Rosetta Earth flybys doesn`t mean the formula is describing a real physical effect. The coincidence could be only accidental and the anomalies are due to failure of the old ODP and the null results of the recent flybys are due to meanwhile corrected ODP. The Earth flyby of Juno on October 9, 2013 could decide, whether the flyby anomaly is real, or not, because the orbit parameters are very similar to those of the first Earth flyby of Galileo with a distinct anomaly value of 3.92mm/s (Table 2). If an anomaly will be detected for the Juno flyby, new physics would be involved without any reasonable doubt. For this case, Paramos and Hechenblaikner [18] proposed an enhanced scientific program with the future STE-QUEST mission for collecting better acceleration data in a highly excentrical orbit. A valuable hint for this mission could be the decision whether such effect would be better described by the Busack or the Anderson equation. According to the Busack simulation (equation (1)) with parameter set used for the prediction of all recent flybys (Table 1), an anomaly of - ( 7.0 +/-2.0) mm/s is expected. The Anderson formula (equation (2)) yields a value of about +6.0mm/s.

## 2  Background of the simulation

Justification for the used formula was given in [4]. For reference purposes this formula is reproduced here in the finally used form.

$$\vec{g}(\vec{r}) = -\frac{G \cdot M \cdot \vec{r}}{r^3}\left[1 + A \cdot \exp\left(-\frac{r - R}{B - C\dfrac{\vec{r} \cdot \vec{v}}{r \cdot v_{Sun}}}\right)\right] \quad (1)$$

with $r = |\vec{r}| \geq R$ and

$G$: gravitation constant  
$M$: field mass  
$R$: radius of the field mass body  
$\vec{r}$: position vector of the test mass with origin at the center of the field mass body  
$\vec{v}$: velocity vector of the field mass center in the gravitational rest frame  
$v_{Sun}$: magnitude of the Sun velocity in the gravitational rest frame  
$A,B,C$: arbitrary constants  

For clarity purposes this could be transformed to

$$\vec{g}(\vec{r}) = -\frac{G \cdot M \cdot \vec{r}}{r^3}\left[1 + A \cdot \exp\left(-\frac{h}{B\left(1 - C'\dfrac{\vec{r} \cdot \vec{v}}{r \cdot v_{Sun}}\right)}\right)\right] \quad (1a)$$

with h = altitude and C'=C/B.

The free parameter A is the amount of the anomalous acceleration at h=0, the parameter B defines the slope of the decrease with altitude, and C'=C/B defines the amount of asymmetry in direction of motion within the assumed gravitational rest frame. The asymmetry introduced by this term is approximately of harmonic first order in direction of the apex of motion in the rest frame.

The consequence of this equation, especially the question why this anomaly wasn't detected yet by analyzing the orbits of Earth satellites, is discussed in detail in [4].

The parameters used for the final evaluation of the Earth flybys of all investigated spacecraft and for the Messenger Mercury flybys as well, were:

```
              low         high
A          : 0.0002453   0.0002306
B / m      :   394000      464000
C'         :   0.3452      0.2263
Vsun / m/s :   360000      360000
RA / h     :   17.78       17.7        | apex
DEC / degree: -37.5        -61         | direction
```

**Table 1**: parameter sets used for all simulations since 2007

The empirical Anderson formula [7] for the anomalous velocity difference $\Delta V$ is

$$\Delta V = 3.099\text{E-}6 * V_\infty * (\cos(\text{dec}_i) - \cos(\text{dec}_o)) \tag{2}$$

with hyperbolic excess velocity $V_\infty$, incoming declination $\text{dec}_i$ and outgoing declination $\text{dec}_o$.

## 3    Simulation of the Juno Earth flyby

In order to derive a measure for the flyby anomaly, the track of a test mass is calculated both for an idealized symmetrical gravity field of the Earth and for a gravity field with additional asymmetry (equation (1)). As shown in [4], the assumption of an idealized gravity field, that means neglecting of the higher order gravity coefficients of the Earth field and of other perturbing gravity sources like Sun or Moon, does not lead to noticeably decreased accuracy. The difference of the velocity magnitudes of the test masses for adjacent positions on the respective tracks was chosen as anomaly deltaV, the value at 5 hours after CA was defined as single measure $\Delta V$ for the flyby anomaly, comparable to the values in the literature. The simulation procedure is described in detail in [4]. The orbit data of the spacecraft were taken from the NASA horizons web interface [5] and reproduced for the test masses with the simulation software.

The free parameters in equation (1) were optimized in [4] for the first 3 flybys with reliable measuring results (Galileo1, NEAR, Rosetta1), with look at additional two flybys with poorer accuracy as known to this time (Cassini, Messenger). Later corrections of some flyby results [1,2,7] were not yet taken into account. Two parameter sets were found with nearly the same small residuals of the flyby anomaly value, but with different curve shapes, i.e. with different "overshoot" in the vicinity of CA. The parameter sets were named "high" and "low" for high and low overshoot (Table1). Such overshoot was seen in the Doppler data of Galileo1, the only flyby with measurements near CA. Unfortunately, the time resolution of the data were poor, so it could not be decided, whether the "low" or the "high" parameter set was the best, so both parameter sets were used to predict following flyby anomalies.

For this article, the simulation software described in [4,6,16,17] was extended by the orbit parameters of the Juno spacecraft as retrieved from the NASA Horizons web interface [5].

The following Table 2 shows the perigee altitude h, perigee velocity $V_p$, hyperbolic excess velocity $V_\infty$, incoming declination $dec_i$ and outgoing declination $dec_o$ for the forthcoming Earth flyby of Juno, compared to the first Earth flyby of Galileo.

|  |  | Galileo1 | Juno |
|---|---|---|---|
| h / km | : | 960 | 559 |
| $V_p$ / km/s | : | 13.74 | 14.60 |
| $V_\infty$ / km/s | : | 8.95 | 9.91 |
| $dec_i$ / ° | : | 12.5 | -14.2 |
| $dec_o$ / ° | : | -34.2 | 39.4 |

**Table 2**: Comparison of the Earth flyby parameters for Galileo1 and Juno

The anomaly curves were calculated using the parameter sets "low" and "high". The results for the flyby anomaly value at 5h after CA are

Parameter set "low":    $\Delta V = -7.72$ mm/s

Parameter set "high":    $\Delta V = -6.23$ mm/s

Interestingly, even with this clear negative flyby anomaly value at 5h past CA the curve of the anomaly shows a positive overshoot in the vicinity of CA. This is shown in Fig.1. Unfortunately, according to the simulation, the vicinity of CA cannot be measured by the stations Madrid, Goldstone and Canberra of the Deep Space Network for tracking spacecraft. Goldstone will lose contact at about 1h before CA while Madrid will acquire contact at about 2h after CA. So, in case no other stations can be activated for tracking Juno, the time range from about -60min to +120min from CA cannot be tracked and the Galileo1 Earth flyby will remain the only one with measured values near CA.

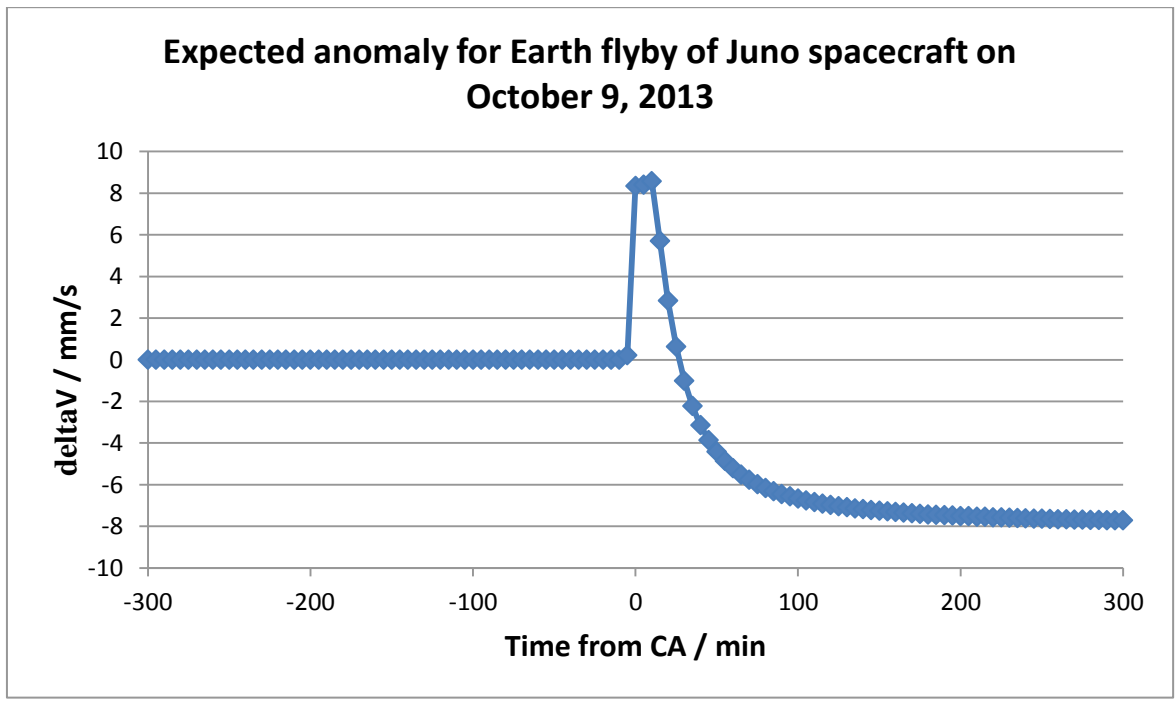

**Fig. 1** : Anomaly curve for the simulation with parameter set "low". The simulation with parameter set "high" is quite similar.

Of course, a well-founded error analysis is impossible for such simulation. Instead, based on the calculations with different parameters the following estimation for the expected anomaly for the Earth flyby of Juno is made.

$\Delta v_{Busack}$ = - (7.0 +/- 2.0) mm/s

With the orbit parameters of Juno from Table 2, the Anderson formula (2) yields without error bar

$\Delta v_{Anderson}$ = +6.05 mm/s

For reference purpose, the simulation software is available as executable program file from [19]. The source code can be requested from the author.

# 6   Conclusion

In this article, the orbit data of the Juno flyby from the HORIZONS web interface were used to compute the expected anomaly for the forthcoming Earth flyby of Juno spacecraft on October 9, 2013 by means of a simulation with the empirical Busack equation [4] and with the likewise empirical Anderson formula [7]. This flyby is an excellent opportunity to decide, whether the flyby anomaly is real, or not, since the orbit parameters are very similar to the Galileo1 flyby with a measured anomaly of 3.92 mm/s. Moreover, in case a distinct anomaly will be seen, it would be clear, whether this effect is better described by the Busack formula or by the Anderson and similar ones, since they predict an anomaly of nearly equal magnitude but with different sign.

The predictions are

Busack:      $\Delta V$ = - (7 +/- 2) mm/s

Anderson:    $\Delta V$ = + 6.05 mm/s   (without error bar)

For this simulation, the same parameter set was used as in [4] and [16], regardless the fact, that the results of some flybys were corrected since 2006 [1,2,7].

The absence of an anomalous effect would support the notion that some failure of the older Orbit Determination Programs was corrected in the meantime, likewise responsible for the null results of the recent Rosetta flybys. The confirmation of an effect would indicate without any reasonable doubt that new physics would be involved, forcing the need of extension of the current standard theory.

To support this, the scientific program of the future STE-QUEST mission could be modified, as suggested by [18].